\documentclass[final, 5p, times, onecolumn]{elsarticle}

\usepackage{amssymb, amsmath}
\usepackage{textcomp}
\usepackage{subfigure}
\usepackage{braket}
\usepackage{float}
\usepackage{xcolor}
\usepackage{cleveref}
\usepackage{multirow}
\usepackage{icomma}
\usepackage{stackrel}
\usepackage{graphicx}
\usepackage{bm}

\begin{document}

\title{Urban Scaling is hardwired in the human brain}

\author{Airton Deppman}
 \address{Instituto de Física -  Universidade de São Paulo, Rua do Matão 1371, São Paulo 05508-090, Brazil
 }
\ead{deppman@usp.br}

\begin{abstract}
The emerging field of the Science of Cities has revealed universal power-law trends in urban scaling, transcending cultural and geographic variations. This study investigates the interplay between fundamental allometry, fractal dimensions, and city social dynamics. By linking these factors to the hierarchical social structures defined by Dunbar’s numbers, a causal relationship is established that illuminates the drivers of urban scaling. Notably, this research highlights the pivotal role of the Big Five personality traits in shaping the fractal nature of urban environments. These findings bridge the gap between human cognitive traits and urban structural patterns, offering a novel perspective on the universality of urban scaling and opening new opportunities for interdisciplinary exploration.
\end{abstract}

\maketitle

Urban scaling is a universal characteristic of cities, transcending their unique socioeconomic and cultural differences. This phenomenon is evident in the power-law behaviour observed in various urban features relative to population size. The universality of urban scaling was empirically discovered through analysis of extensive data provided by social media platforms and mobile devices~\cite{Batty2013}. Contrary to earlier notions of chaotic urban behaviour, populations demonstrate consistent organizational patterns around urban centres. While these universal patterns do not capture all dimensions of urban life, they coexist with the distinctive qualities that define each city worldwide.

A city cannot be fully represented by deterministic equations, as universal laws provide only rough approximations of urban phenomena. Individual cities show significant deviations from these trends, underscoring their unique characteristics. Nevertheless, these universal predictions help distinguish which aspects of urban life are shaped by specific local attributes. For those features deemed universal, their ubiquity cannot be attributed to geographic, cultural, or historical factors due to the wide diversity of these aspects across cities globally. Instead, the origins of these universal aspects can be traced to the one common factor in urban formation: the human being.

This study focuses exclusively on the universal features of urban life, seeking to uncover the underlying causes of the common characteristics shared by cities. It examines objective trends that emerge spontaneously across networks of urban areas, identifying causal relationships among different aspects of social behaviour associated with the power-law patterns observed in universal urban characteristics. The analysis then extends to explore which human attributes might drive the universality seen in the data. This approach combines mathematical models, social dynamics, and anthropological insights, exemplifying the convergence of mathematical and social sciences in advancing our understanding of human nature.

Simple power laws, rather than complex cultural factors, describe how various socioeconomic outputs of urban centres scale with population size. Superlinear or sublinear trends are frequently observed~\cite{Ribeiro2023}, challenging the conventional practice of comparing per capita outputs, as is common in the assessment of a country’s gross domestic product, for example. The development of theories incorporating complex system dynamics has provided tools for exploring urban scaling~\cite{Chen2023}. It has become clear that allometric relationships contribute to the emergence of power-law trends, although these models also require the inclusion of a fractal dimension~\cite{Bettencourt2007}. Recall that fractals are systems presenting a fine internal structure and invariance under scale transformation~\cite{Falconer2013-xt}, leading to self-similarity and geometry with fractional dimension.

The fundamental allometric relationship, linking urban infrastructure area ($A$) to population size ($N$), is expressed as\footnote{The allometric relation connects the urban area to population size and is inspired by the metabolic scaling relationships observed in mammals.}
\begin{equation}
 A = a_o N^{\beta} \,, \label{eq:fundamentalallometry}
\end{equation}
where $a_o$ represents the baseline area and $\beta$ is the scaling exponent. The findings highlight that $\beta$ is a fractional value, deviating from the simple expectation that total urban area would increase linearly with population size, which would correspond to $\beta = 1$. In this context, universality means the $\beta$ is approximately constant for cities worldwide~\cite{Batty2013,Ribeiro2023,Bettencourt2007}. The multiplicative parameter $a_o$, however, can vary according to specific characteristics (e.g., household size) of the regions where the cities are located.

The fractal geometry\footnote{Fractals are mathematical or natural structures characterized by scaling invariance and fine internal details. In geometry, these properties often result in spaces with fractional dimensions. A key feature of fractals is self-similarity, meaning that a portion of the structure resembles the entire structure when magnified.} of urban areas can be observed through cartographic analyses. The box-counting technique accurately determines the fractal dimension ($d_f$) of cities~\cite{Encarnao2012}. If $\ell$ represents the city's typical linear length, its area is given by $A = \ell^{d_f}$. The fractal dimension typically falls within the range $1.4 < d_f < 1.9$, which is smaller than the expected integer dimension $d=2$. Key questions arise: What is the relationship between the scaling exponent and the fractal dimension? Why do humans exhibit fractal-like behaviour?

Recent developments have provided a mathematical framework for describing the dynamics of systems that move through fractal spaces~\cite{MGD2024}. In such systems, dynamics follow a nonlinear equation where the entropic index from Tsallis Statistics~\cite{TsallisBook} plays a crucial role\footnote{Tsallis Statistics is a generalization of Boltzmann Statistics, proposed in 1988 to address the growing number of systems that standard statistical mechanics could not appropriately describe. In particular, this generalized framework is especially effective in modelling complex systems.}. When applied to model population density dynamics in cities, this framework yields the following relationship~\cite{DEPPMAN2025115877}: 
\begin{equation}
\beta = 1 - \frac{d}{2}(q-1) = 1 - \frac{d- d_f}{2} \,.\label{eq:betaqdf}
\end{equation}
These equations establish connections between $\beta$, $q$, and $\delta d_f=d-d_f$, where $d_f$ is the fractal dimension of a fractal space embedded in a space with integer dimension $d$. The entropic index, $q$, is a parameter in the nonlinear equation describing anomalous diffusion in fractal spaces, and in the nonextensive statistics as a measure of entropy non-additiveness, with the standard statistics being recovered for $q=1$\footnote{Eq.~(\ref{eq:betaqdf}) is derived by analyzing the dynamics of a population within an urban area. It employs a nonlinear form of the Fokker-Planck equation, which is effective for addressing nonextensive systems and, in particular, diffusion in fractal spaces.}.Eq.~(\ref{eq:betaqdf}) was validated through an analysis of 319 cities.

Note that each parameter in Eq.~(\ref{eq:betaqdf}) can be independently measured using urban life data. The fractal dimension is determined through fractal analysis of the city's satellite images, while the parameter $q$ is measured by examining the population distribution within the city. The scaling exponent is obtained by comparing the growth in area with population size across a set of cities. Although Eq.~(\ref{eq:betaqdf}) establishes the relationship between $\beta$, $d_f$, and $q$, addressing the first question posed above, further investigation is needed to understand the causal connections among the processes related to fundamental allometry, the fractal dimension, and the entropic index before moving to the second question.

In the following steps, this work examines causal relationships and proposes a conjecture regarding the mechanisms leading to the formation of fractal urban areas. It suggests that two out of the three potential origins of observed urban scaling are more likely to be consequences rather than causes of fractal organization, thus highlighting the third as the most probable cause. Furthermore, the analysis draws connections between the neurological properties of the human brain and the universal drivers of urban scaling, providing additional support for the causal relationships among the various factors shaping urban organization.

A city fundamentally operates as a mechanism for generating social outputs, a concept central to various urban scaling models~\cite{Ribeiro2023, Bettencourt2007}. Recent analyses reinforce this perspective by identifying links between urban scaling and the percolation fractal exponent~\cite{DEPPMAN2025115877}. In percolation systems, a critical density of connections between nodes determines when the system becomes fully interconnected, spanning from one end to the other\footnote{In the allometric models, a city's infrastructure, similar to the circulatory system of mammals, is built to ensure that all individuals have access to any place within the city, forming a percolative system.}. When applied to cities, this suggests that once a critical threshold of connectivity is achieved, individuals can travel easily across the city. Consequently, cities tend to evolve toward this critical percolation threshold, undergoing an organic development process to enhance connectivity among inhabitants. However, there is no indication that individuals consciously navigate fractal paths or adjust their movements to produce a power-law output. In this context, the scaling exponent emerges as a byproduct of urban organization, shaped by the population and infrastructure dynamics designed to maximize social output. This implies that urban output scaling is not the primary cause of fractal characteristics in city structure; rather, population dynamics and fractal geometry likely play more fundamental roles in driving these patterns.

Urban planning traditionally relies on a rational approach to designing infrastructure typically based on Euclidean geometry. In this context, fractal aspects are not initially considered. Moreover, in many cases, the actual configuration of urban infrastructure results from centuries of continuous development and organic growth, where calculated planning has played a limited role. Consequently, fractal geometry likely arises more from social needs than from rational design decisions. This suggests that the dynamic flow, characterized by the variable $q$, determines the fractal geometry with dimension, $d_f$, clarifying the causality relation in Eq.~\ref{eq:betaqdf}. These insights reveal that human behaviour is resilient enough to transcend Euclidean geometric constraints, fostering the emergence of fractal geometry in the organic development of cities.

The above analysis highlights some of the key drivers contributing to the fractal geometry of cities. Population distribution within an urban area stems from a nonlinear dynamic process, as described by the PPE. The infrastructure is designed to accommodate these population dynamics, resulting in a fractal pattern that shapes the city’s geometry. This fractal structure optimizes the movement of people, thereby enhancing social output. Consequently, allometric scaling arises as a byproduct of this efficient fractal configuration. The primary drivers of this structure are the dynamic processes characterized by the parameter $q$. Further investigation is needed to understand the role and significance of this parameter in the context of social life.

Studies on information spreading and the epidemic evolution of diseases~\cite{Deppman2021,Policarpo2023} highlight the significance of social contact in information diffusion and social output production. These models assume a fractal structure of social contacts, which leads to nonlinear information diffusion and results in informational distributions with a nonextensive nature. The models indicate that the dynamical parameter $q$ is determined by the average number of close contacts ($n_c$) through the relation $q-1=1/n_c$\footnote{This relation was validated by an analysis of COVID-19 spread presented in Ref.~\cite{Policarpo2023}}.
Using the relation between $q$ and $n_c$ along with  Eqs~\ref{eq:fundamentalallometry} and~\ref{eq:betaqdf}, it follows that
\begin{equation}
A = \left(\frac{a_0}{N^{1/n_c}} \right) N \,.
\end{equation}
This equation shows that individuals distribute themselves efficiently, sharing space with close contacts. Rather than a fixed baseline area $a_0$, the effective area per individual becomes $a_0/N^{1/n_c}$, highlighting how social dynamics shape urban geometry. Individuals accept sharing their space with a few close contacts to maximize the social output of their socioeconomic activities. As a result, the total area occupied by the social groups is non-additive.

The model proposed in Ref~\cite{Policarpo2023} aligns with findings from sociological and anthropological studies, which have observed an approximately self-similar pattern in the social interactions of primates~\cite{Zhou2005}. These studies have shown that the number of members in the support clique, i.e. the size of the close contact group, is related to the size of the primate's neocortex. This suggests a strong relationship between brain structure and the number of close social connections, with larger neocortex sizes allowing for more close contacts, in what has been called the social brain hypothesis~\cite{Byrne1988-rl,Dunbar2007}. For humans, the support clique size was estimated around 3-5 individuals.

The fractal model of information spread assumes a hierarchical, self-similar structure for social interactions, where individuals belong to a close contact group consisting of approximately $n_c \sim 5-7$ people~\cite{Policarpo2023}. Self-similarity implies that groups at one hierarchical level interact with a fixed number of similar groups, equal to $n_c$. This pattern repeats across different levels of the hierarchy, maintaining the same interaction structure throughout.  This aligns with recent studies indicating a range of $4 \leq n_c \leq 10$ for individuals in urban areas globally~\cite{Ghafoor2019}. These results are remarkably similar to the social brain results.

The above considerations emphasize the critical role of the number of contacts in social interactions. This number, linked to the hierarchical structure of social behaviour, governs the flow of information within society by determining the nonlinearity of the social dynamics regulated by the parameter $q$ in the equation describing the dynamic process. Consequently, this study clarifies the relationships among key social aspects: the number of close contacts, social dynamics within a hierarchical structure, fractal geometry, and the scaling of social outputs. It explores the causal relationships underlying urban processes, suggesting that individual behaviour —particularly the number of close contacts — serves as the primary driver of fractal urban geometry. Yet, one fundamental question remains: why is such behaviour universal?

Indeed, the observed data shows a nearly constant scaling exponent worldwide. Using Eq.~(\ref{eq:betaqdf}), this suggests a constant value for $q$ across cities, and therefore, a consistent number of contacts per individual. This constancy aligns with the idea that social dynamics shape the fractal geometry of urban areas. Thus, the question can be reframed as: why do humans prefer a limited number of contacts? Specifically, why does an individual in a large city maintain a similar number of close contacts as someone in a smaller city?

The answer lies in the human brain and its evolutionary process. Fast cognition of social traits in other individuals may have played a central role in the survival of Homo Sapiens~\cite{Hamilton1980}. The human brain, therefore, is designed to allow for a fast mechanism of interpretation of relevant traits. This is especially important for the first impressions of unknown individuals, which are mostly related to the interpretation of facial traits.

Functional magnetic resonance imaging (fMRI) studies reveal how social recognition is formed in the human brain~\cite{Hamilton1980}. It has been shown that attractiveness and trustworthiness are the social characteristics that are more immediately recognized because they are important for the preservation of the species and survival. Specific brain regions are activated for identifying others, and it is even possible to determine from fMRI who a subject is thinking about~\cite{Hassabis2013}, proving that personality traits are registered in the brain by creating a personality model. These models are essential for a successful social life.

Concerning facial recognition, studies have shown that a few characteristics are modelled in the brain to make an identification of close others or strangers. The process involves a vectorial combination of a few facial traits that are used by the brain for fast identification of individuals in a mechanism called eigenfaces~\cite{Todorov2008}. %~\cite{Todorov2008, Todorov2009}
This mechanism has been adopted for fast identification of faces by artificial intelligence~\cite{Turk1991}.

The accepted consensus suggests that the crucial personality traits influencing social interactions are five in number~\cite{Goldberg1990}. These fundamental traits, collectively referred to as the Big Five, are believed to encompass the essential social characteristics that individuals must evaluate to maintain a healthy social life\footnote{The Big Five personality traits are: Openness, Conscientiousness, Extraversion, Agreeableness and Neuroticism.}. %\cite{Goldberg1990, Gosling2003, Roccas2002}
While these traits do exert some influence on the number of friends an individual may have, the observed variations are relatively minor\cite{Kang2023}. The average number of close friends aligns with findings from other independent studies. It is reasonable to posit that, given the necessity to identify a limited number of traits, the human brain specializes in learning a minimal set of personality models rather than generating numerous models. Consequently, for fostering a robust social life, it is preferable to intimately know a select few close contacts (such as family and friends) rather than maintaining superficial connections with a large number of acquaintances.

The findings support the social brain hypothesis, highlighting the deep links between human brain structure and social behaviour, particularly the number of close relationships an individual can maintain\footnote{The hierarchical social structure was proposed as an extension of Dunbar's number, which represents the limit on the number of friends a person can maintain—approximately 150. Further analysis by Dunbar and colleagues revealed that the group of 150 friends exhibits non-uniformities in the types of social interactions, leading to a hierarchical structure that is approximately scale-free, similar to a fractal structure.}. Studies of social interactions suggest a fractal organization in human social structures, with individuals forming layered networks based on social closeness. This hierarchical, fractal pattern in social organization constrains how people move within urban spaces: individuals preferentially visit locations where they can encounter familiar faces and often travel with acquaintances. Since commuting behaviour plays a key role in urban planning decisions, the fractal nature of social relationships significantly shapes the design and layout of cities. Consequently, urban areas often exhibit fractal geometry, and social dynamics within these spaces follow a power-law distribution.

The assumption presented above can be subjected to objective, quantitative tests. Most studies begin by employing personality models, such as the Big Five, to evaluate the number of close social contacts~\cite{Kang2023}. Future research, however, could take an inverse approach, starting with the analysis of close contacts to investigate how these relationships span the five-dimensional space of personality traits. Modern technologies now enable a detailed examination of how social connections influence the spread of information and the movement of individuals within cities~\cite{Wang2012, Senaratne2018}, providing valuable insights into the hierarchical structure of social environments.

In conclusion, this work reviews recent advances in understanding the dynamics of urban life. While the framework in Ref.~\cite{DEPPMAN2025115877} focuses on the physical and mathematical modelling of urban scaling, this study bridges these findings with the neurological and cognitive mechanisms that underpin human social behaviour, providing a deeper understanding of the universal drivers of urban scaling.
The anomalous diffusion of the population through fractal urban spaces establishes a connection between the fractal dimension and the scaling exponent. This anomalous diffusion is characterized by a non-linear equation associated with fractal geometries and non-extensive statistics. Notably, the entropic index, $q$, serves as a key parameter governing anomalous diffusion, highlighting the relevance of Tsallis statistics in bridging natural and social sciences~\cite{Tsallis2023}.

The present analysis reveals causal connections among the scaling exponent, the fractal dimension, and the entropic index, demonstrating how these diverse aspects of urban dynamics emerge. It is found that population dynamics represent the fundamental factor influencing the other two aspects. Furthermore, the association of the entropic index with the number of close social contacts suggests that this metric plays a pivotal role in determining urban dynamics. The universality of fundamental allometric relationships is attributed to the relative independence of the number of close contacts from the geographic region in which an individual resides.

Further considerations reveal that a characteristic of the human brain, namely the five dimensions of personality traits, shaped during the evolutionary process, controls the number of close contacts, influencing social behaviour, and ultimately determining the infrastructure design in urban areas.

\bibliographystyle{ieeetr}
%\bibliography{OriginUrbanScaling.bib}

\end{document}